# BreathGRU: A Novel Semi-Supervised Bidirectional Gated Recurrent Unit Framework for Speech and Breath Segmentation for Respiratory Audio

Sania Fatima Sayed[1], John W. Holloway[2], Reyer Zwiggelaar[1], Faisal I. Rezwan[1,2,*]

[1]Department of Computer Science, Aberystwyth University, Aberystwyth, United Kingdom, UK SY23 3DB
[2]Human Development and Health, Faculty of Medicine, University of Southampton, United Kingdom, SO16 6YD

Corresponding Author: Faisal I. Rezwan (far8@aber.ac.uk)

## Abstract:

Speech-breath segmentation is a fundamental preprocessing step in respiratory audio analysis, enabling applications such as respiratory acoustic biomarker extraction, lung function prediction and disease monitoring. Existing approaches, including threshold methods, Fourier Transform-based techniques, and unsupervised and pretrained voice activity detection (VAD) models, primarily focus on speech detection and often classify breathing events as non-speech or silence, limiting their applicability for precise breath detection. To address this limitation, we propose BreathGRU, a semi-supervised Bidirectional Gated Recurrent Unit (BiGRU) framework specifically designed for speech-breath segmentation. The proposed framework combines frame-level acoustic feature extraction with bidirectional recurrent modelling, pseudo-label refinement and duration-constrained Segmental Viterbi decoding to produce speech and breath segmentation. BreathGRU was evaluated against the existing approaches, using manually annotated recordings. Performance was assessed using event-based, time-based, overlap-based, duration-based and boundary-based segmentation metrics. Experiment results demonstrated that BreathGRU achieved the highest breath event recall (0.83), the lowest onset-localisation error (0.14s) and the highest Mean Match Intersection over Union (0.81), with competitive overall segmentation performance compared to large pretrained VAD models like Silero. Qualitative evaluation on manually annotated recordings further showed close agreement between BreathGRU and manual annotation, with better breath detection compared to Silero. Additional evaluation on external Coswara and publicly available recordings demonstrates consistent segmentation behaviour with diverse recording conditions and speaker accents. These findings demonstrate that explicit breath event modelling provides advantages over general-purpose VAD models and establish BreathGRU as an effective speech-breath segmentation framework which can be applied for respiratory audio analysis and pulmonary healthcare applications.

## 1. Introduction

Audio analysis has become an increasingly important area of research in healthcare, where speech, voice, phonetic and respiratory sounds are utilised as non-invasive biomarkers for pulmonary disease assessment [1], [2], [3], [4]. In the case of speech analysis, accurate speech-breath segmentation is a fundamental preprocessing step in audio analysis applications [5]. Errors during segmentation directly affect the quality of extracted features and, subsequently, the performance of predictive models. Hence, robust and accurate speech-breath segmentation is a critical component of any audio analysis pipeline.

Speech and breath sounds exhibit significant acoustic differences and exhibit distinct temporal, spectral, fractal, and cepstral characteristics [5]. Speech signals demonstrate higher energy and more spectral content, whereas breath sounds show lower energy, broader spectral content, and different spectral patterns [6], [7], [8], [9] (Figure 1).

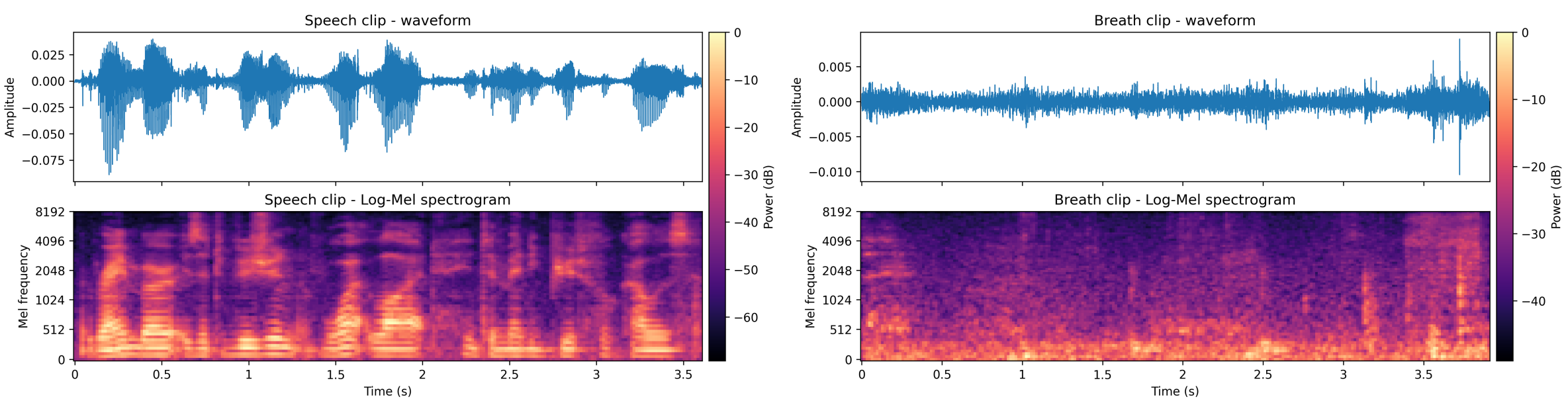

***Figure** 1: Representative frequency and Mel-Spectrogram from speech and breath clips*

A variety of speech-breath segmentation approaches have been proposed [5], [7], [8], [9], [10], [11]. Traditional methods rely on handcrafted acoustic features and threshold-based decision rules [10]. Signal processing approaches utilise frequency-domain representations and spectral masking techniques to distinguish between speech and non-speech regions [9], [12]. More recently, machine learning and deep learning-based approaches have been investigated, including autoencoder-based frameworks and Voice Activity Detection (VAD) systems, such as Silero [11] and PyAnnote [13] have become increasingly popular due to their strong speech activity detection performance.

Respiratory audio has been increasingly investigated for pulmonary health monitoring. Previous studies have analysed respiratory audio recordings and investigated approaches for disease monitoring, classification, and spirometry prediction [1], [2], [3], [4], [14], [15], [16], [17]. Existing work has benchmarked CNN-LSTM [4] architectures for spirometry prediction and LTSD-based VAD methods for detecting speech, pauses, and breathing signals [14]. However, the majority of existing methodologies have focused on speech activity detection or utilised phonemic and vowel sounds [18] rather than explicitly modelling breath events. Consequently, many existing segmentation approaches were not specifically developed for speech-breath segmentation [19], [20].

A rule-based threshold method was developed by Alam et al. [21] for segmenting speech and breath regions. These methods often rely on manually selected thresholds that are sensitive to recording conditions. In Alam's model, a set of energy-envelope and spectral-based features was extracted from both segments. Segments that did not satisfy these conditions were classified as speech. While a Fourier transform-based method was proposed in a previous work, which established the use of the short-time Fourier transform (STFT) to separate speech and non-speech regions [12] (*Supp. Table 1* specifies the parameters and values). Ruinskiy and Lavner developed an autoencoder-based approach [19] that uses heuristic post-processing to distinguish speech and breath segments. A convolutional autoencoder architecture was used to learn latent representations of the extracted features (see Supp. Figure 1).

Although the pretrained VAD models, PyAnnote and Silero , achieve excellent speech detection performance, they are primarily optimised to identify speech from non-speech regions. Both of these models often treat inhalation, and exhalation sounds as background or silence rather than clinically meaningful breathing segments. PyAnnote (pyannote.audio) is a deep learning Python toolkit for voice activity detection tasks, speaker diarisation and segmentation. The model architecture is based on neural networks that learn from temporal speech representations from large, annotated speech datasets and give frame-level speech predictions. In contrast, Silero is a lightweight deep learning-based speech detection model trained on a large multilingual speech dataset. The model analyses audio frames and results in a probability estimate for the presence of speech in the given audio. All these existing methods rarely investigate the physiological characteristics of respiratory signals, like temporal continuity, expected breath durations, and speech-breath boundary localisation, that are commonly encountered in respiratory audio recordings.

Hence, to address these limitations in the existing frameworks, we propose BreathGRU, a novel semi-supervised Bidirectional Gated Recurrent Unit (BiGRU) framework specifically designed for speech and breath segmentation. This model is trained using the Recurrent Neural Network (RNN) model of BiGRU on annotated breath and speech segments with extracted features. The model calculates speech and breath probabilities based on temporal estimation.

Moreover, this study performs a comprehensive comparison between speech-breath segmentation approaches spanning rule-based, digital signal processing, unsupervised learning, pretrained VAD systems, and the proposed BreathGRU framework. The methods were evaluated on manually annotated recordings using time-based, event-based, overlap-based, duration-based, and boundary-based metrics. Qualitative comparison was also performed on external recordings obtained from the Coswara database and publicly available YouTube recordings to illustrate the segmentation behaviour of different approaches. The results provide detailed insight into the strengths and limitations of traditional segmentation methods, VADs and the proposed framework.

## 2. Methods

### 2.1. Dataset

The dataset includes 453 speech recordings from 44 patients with clinically diagnosed asthma after bronchial hyperresponsiveness (BHR) testing using methacholine challenge and mannitol. The data were collected from 27 patients (354 recordings) at Newcastle University in Australia and 17 patients (99 recordings) at the University of Southampton (United Kingdom). The recordings included patients reading the Rainbow Text or a paragraph from “A Winter Book” for a minute after dosages of the bronchoconstrictors. All participants gave written informed consent, and the study was approved by the relevant Local Ethics Committee (number 12/EE/0545 in the UK and 2020/ETH01978 in Australia).

The recordings present audible wheeze, heavy inhalation and exhalation while reading the given passage. These audio recordings were manually extracted and annotated as speech and breath segments with timestamps and used as ground truth to evaluate five different speech or breath detection and extraction methods. To evaluate the generalisation of the models, five clips from different patients with slow and fast counting from the Coswara dataset [22] and 60-second clips from five different TED talks [23], [24], [25], [26], [27] were used.

## 2.2. Breath Bidirectional Gated Recurrent Unit (BreathGRU)

BreathGRU is a semi-supervised BiGRU [28] model to perform speech and breath segmentation. Unlike conventional VAD algorithms, which primarily distinguish speech from silence, BreathGRU is designed to identify speech and breath from temporal continuity and physiological characteristics of respiratory sounds. The complete architecture is illustrated in Figure 2. BreathGRU consists of five components: acoustic feature extraction, bidirectional recurrent unit modelling, frame-level classification, semi-supervised training, and duration-constrained decoding.

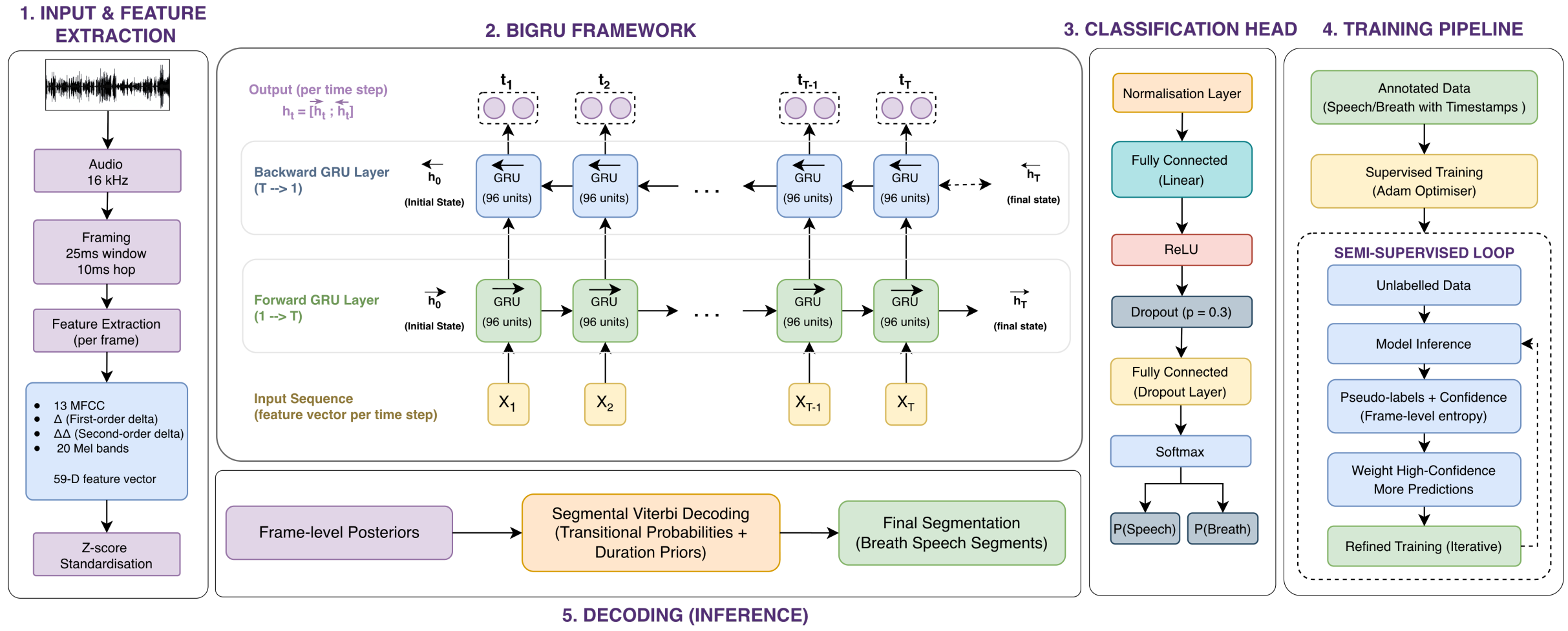

***Figure 2:*** *A detailed view of the BreathGRU architecture. The framework is divided into feature extraction, bidirectional GRU units, classification heads, training pipeline and inference components*

### 2.2.1. Acoustic Feature Extraction

Each recording was initially resampled to 16kHz to provide a uniform sampling frequency across all recordings to preserve the spectral characteristics required for respiratory sound analysis. The audio signal was then segmented into overlapping frames using a 25ms analysis window with a 10ms hop length, a configuration commonly adopted in speech processing because it provides an appropriate balance between temporal and spectral resolution. Each frame representation consists of a 59-dimensional acoustic feature vector capturing complementary spectral and temporal characteristics of speech and breath sounds. Features were extracted and concatenated using the Python librosa and NumPy libraries. The extracted features comprised: 13 Mel-Frequency Cepstral Coefficients (MFCCs), 13 first-order delta coefficients, 13 second-order delta coefficients, and 20 Mel-spectrogram frequency bands.

MFCCs represent the spectral envelope of speech and breath sounds, while the delta ($\Delta MFCC$) and delta-delta ($\Delta^2 MFCC$) coefficients capture short-term temporal dynamics between consecutive frames. The Mel-spectrogram bands preserve detailed frequency energy distributions that are often informative for distinguishing speech from inhalation and exhalation sounds. Finally, each feature dimension was standardised using Z-score normalisation to reduce variability and to improve optimisation during training the network.

The resulting training vector for frame $\boldsymbol{t}$ can be expressed as:

$$x_t = [MFCC, \Delta MFCC, \Delta^2 MFCC, Mel]_t$$

Where $\boldsymbol{x_t}$ denotes the acoustic representation supplied to the segmentation model.

### 2.2.2. Bidirectional GRU Network

The extracted acoustic features were processed using a two-layer BiGRU network, as these are suitable for sequential respiratory recordings. This is because the classification of an individual frame depends not only on the local acoustic properties but also on its surrounding temporal context. The BiGRU is particularly used to simultaneously analyse the feature sequence in both preceding and succeeding frames. The forward GRU captures the dependencies from the preceding frames, while the backward GRU incorporates future contextual information. This enables more reliable identification of speech and breath events occurring over transitional boundaries.

For each time frame $\boldsymbol{t}$, the hidden representations are computed:

$$\overrightarrow{h_t} = GRU_f(x_t, \overrightarrow{h_{t-1}})$$

$$\overleftarrow{h_t} = GRU_b(x_t, \overleftarrow{h_{t-1}})$$

where $\boldsymbol{GRU_f}$ and $\boldsymbol{GRU_b}$ denote the forward and backward recurrent layers

The final hidden representation is obtained by concatenating both directions, which provides a complete contextual representation for each acoustic frame:

$$h_t = [\overrightarrow{h_t}; \overleftarrow{h_t}]$$

The proposed network consists of two stacked bidirectional layers, each with 96 hidden units per direction. Stacking multiple recurrent units enables higher-level temporal representations to be learned while allowing the model to capture both short-duration acoustic events and longer respiratory patterns.

#### 2.2.3. Frame-level Classification Head

The contextual representations generated by the BiGRU network were passed through a lightweight classification head to estimate frame-level posterior probabilities for speech and breath classes. Layer Normalisation was applied to stabilise the hidden feature distribution and improve optimisation, and the normalised features were projected through a fully connected layer and a ReLU activation function to enhance feature discrimination.

To reduce overfitting, a dropout layer (p = 0.3) was added before the final output layer. The final layer consists of a two-node fully connected layer representing probabilities of speech and breath. These probabilities were computed by a Softmax activation function.

$$P(y_t|x_t) = Softmax(Wh_t + b)$$

where $\boldsymbol{P(y_t|x_t)}$ denotes the probability of the two classes – speech and breath for frame $\boldsymbol{t}$. $\boldsymbol{W}$ denotes the weight matrix of the fully connected layer, $\boldsymbol{h_t}$ denotes the contextual feature set generated by BiGRU, and $\boldsymbol{b}$ denotes the bias vector.

#### 2.2.4. Semi-Supervised Learning Strategy

Annotating speech and breath boundaries is an intensive process. Therefore, to overcome this limitation, the BreathGRU model uses a semi-supervised learning strategy that utilises both labelled and unlabelled recordings. The annotated speech and breath segment labels were used to train the initial BiGRU model. After convergence, the trained model was applied to the unlabelled recordings to generate pseudo-labels. Since the pseudo-label quality varies between the recordings, prediction confidence was estimated using frame-level entropy

$$H(p) = -\sum p_i \log p_i$$

Where $\boldsymbol{p_i}$ denotes the predicted probability of class (speech or breath) $\boldsymbol{i}$ for the given frame. Entropy $\boldsymbol{H(p)}$ measures the uncertainty of the model's prediction for each frame.

Here, lower entropy corresponds to higher prediction confidence. Frames with lower entropy (higher confidence) were assigned greater importance during training iterations, whereas uncertain predictions contributed less to the parameter updates. The confidence-weighted pseudo-labelling refined the segmentation model while reducing the influence of unreliable predictions. Consequently, the proposed BreathGRU framework effectively used large quantities of unlabelled respiratory recordings without requiring manual annotation.

#### 2.2.5. Duration-Constrained Viterbi Decoding

Although the classification network estimates the frame-level posterior probabilities, the respiratory/breath events show strong temporal continuity and hence should not fluctuate rapidly between speech and breath events. To improve temporal consistency, log-normal duration distributions were computed for speech and breath segments using annotated training recordings. These duration priors capture the expected length of each event in the training data and provide additional temporal constraints during decoding.

The final segmentation sequence was obtained using Segmental Viterbi Decoding, which determines the optimal sequence of speech and breath states by jointly considering frame-level posterior probabilities, state transition probabilities and duration priors. Rather than selecting the most probable class for each frame, the decoder identifies the sequence that maximises the overall likelihood with respiratory durations. Rather than predicting each frame independently, the decoder considers the entire sequence with expected durations of speech and breath segments. This reduces short, fragmented predictions and produces smoother segmentation boundaries that better reflect the natural speech-breath patterns.

### 2.3 Evaluation Metrics

The performance of the speech-breath segmentation models was evaluated using the rule-based threshold method, Short-Time Fourier Transform (STFT), Unsupervised autoencoder, pretrained VAD models (Pyannote and Silero), along with the proposed BreathGRU method. These methods were evaluated by comparing the predicted speech and breath segments with manually annotated labels, and to quantify the performance, overlap-based and boundary-based metrics were used. The formulas for calculating each of the metrics are in the Supplementary Material (Formulae 1 – 6).

i. **Event-based F1 Score:**
Event-based F1 score evaluates each correctly identified individual speech and breath segment, and a segment or event was considered correctly classified when it matches the reference segment according to the overlap criteria. Precision, recall and F1Score were calculated using the number of correctly detected events. This focuses on detecting complete speech and breath segments rather than frame-level temporal agreement like the Tim-based F1 Score.

ii. **Time-based F1 Score:**
The Time-based F1 score gives the temporal agreement between the predicted and the reference segment labels for the entire recording. This metric measures how accurately a method classifies the signal over time while accounting for both false positive and false negative regions. Precision and recall were calculated based on the duration of the identified or predicted speech or breath region and combined using harmonic mean.

iii. **Onset and Offset Error:**
These errors are calculated based on boundary localisation accuracy, where onset error is the absolute temporal difference between the start times of the predicted and reference segments, while offset is the difference between the end times of the segments. Lower values of the errors mean more accurate segment boundaries.

iv. **Mean Duration Error:**

Mean Duration Error gives the difference between the predicted segment duration and the reference segment duration. For each matched speech or breath segment, the absolute difference in duration was calculated and averaged across all segments. Lower values indicate greater agreement between the segment lengths.

v. **Mean Match Intersection Over Union (IoU):**
The Intersection over Union (IoU) quantifies the overlap between predicted and reference segments. IoU consists of two different metrics: Time IoU, indicating temporal overlap between the predicted and reference labels and Mean Match IoU, which quantifies how well the matched segments overlap once detected

vi. **Boundary Tolerance:**
Boundary Tolerance measures the proportion of the predicted boundaries that fall within the predefined temporal boundary of the reference segment.

## 3. Results

Evaluation metrics showed peak performance for the proposed BreathGRU and Silero methods. The metrics are computed for each speech and breath.

### 3.1. Event-based Evaluation:

Silero and BreathGRU compete in detecting the two events, F1 scores for Silero are 0.57 and for BreathGRU are 0.52. In contrast, recall scores show identical performance (Silero recall = 0.68, BreathGRU recall = 0.67), signifying both methods detect true events equally (*Figure 3 and Supp. Table 2*). BreathGRU excelled at breath event detection with an F1 and recall of 0.61 and 0.83, respectively, while Silero had marginally lower scores (F1 = 0.55, recall = 0.67). Threshold was competitive (0.51), Pyannote detected fewer events, resulting in low recall (0.27), while Fourier (0.09) and autoencoder (0.09) performed poorly overall. For speech events, Silero showed the strongest F1 scores (F1 = 0.59) as expected from a VAD, while BreathGRU was the second-best model (F1 = 0.43).

### 3.2. Time-based Evaluation:

BreathGRU model showed the best overall recall (0.83) with strong breath segmentation recall scores of 0.95 and speech recall of 0.71. Silero achieved the highest performance (F1 = 0.814) for speech frame detection (speech F1 = 0.94), while breath was moderate (breath F1 = 0.68). Pyannote had the best speech precision but very low breath recall. The threshold method, autoencoder and Fourier achieved moderate overall scores. Meanwhile, the threshold method demonstrated strong breath recall but was ineffective for speech segments. The Fourier method, though it achieved the highest recall for breath, showed significantly low precision, which suggests that the method misclassifies segments as breath. The autoencoder showed poor segmentation performance. For the time-based F1 scores. The detailed scores are shown in *Figure 4 and Supp. Table 3*.

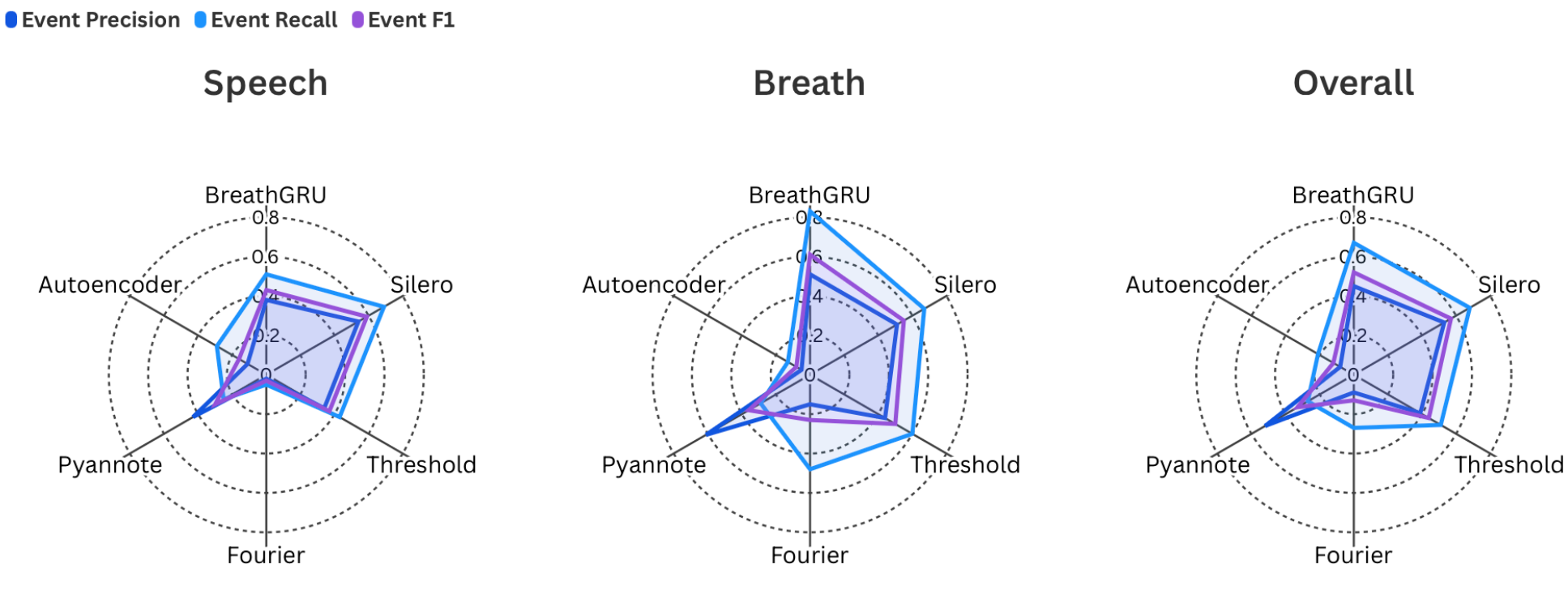

***Figure 3:*** *The radar chart presents mean precision, recall, and F1 scores for Time-based segment detection for speech, breath, and the combination of the two for all the models. BreathGRU was the best-performing model for breath time-based segment detection amongst the six methods.*

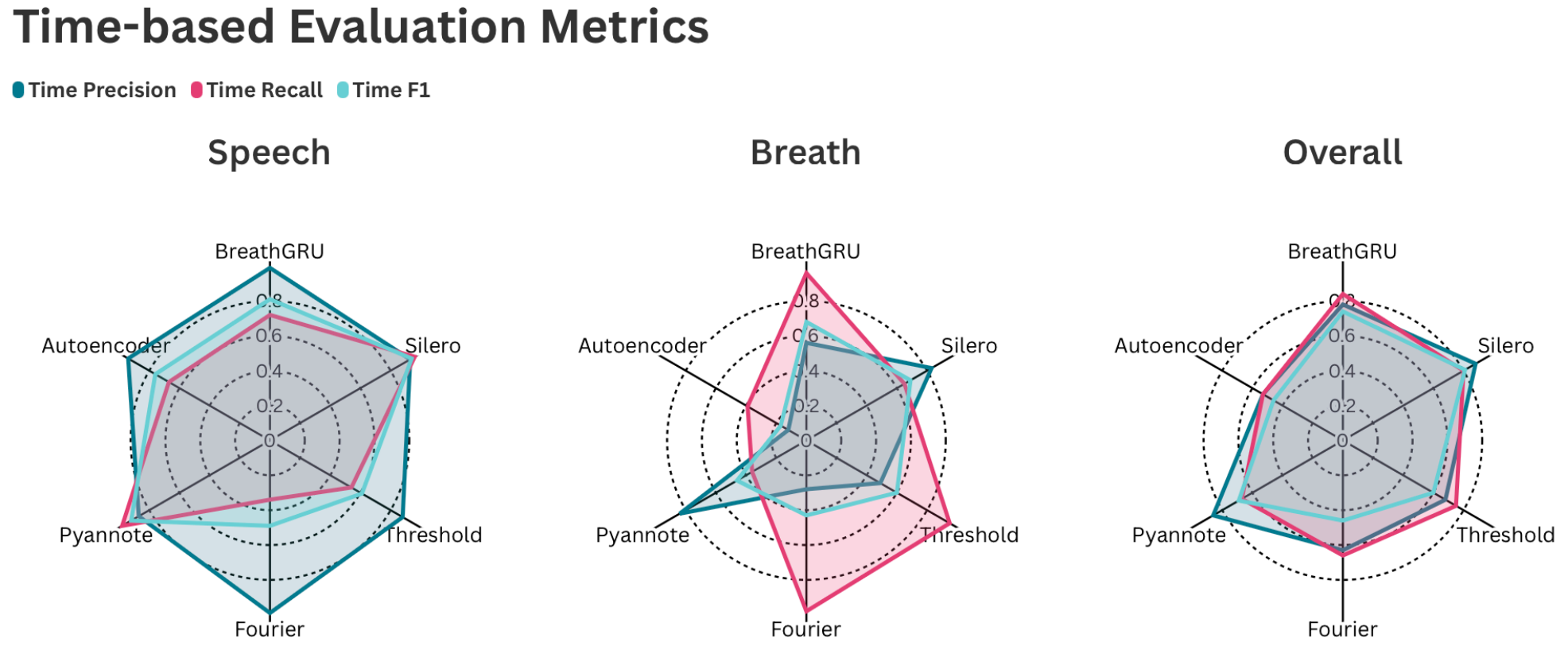

***Figure 4:*** *This radar plot shows Event precision, recall and F1 scores for speech, breath and overall, for each of the six models. BreathGRU is better at breath recall for time-based evaluation metrics.*

### 3.3. Onset and Offset Evaluation:

BreathGRU gave the most accurate localisation of segment start times, given by the mean onset error (0.18s), while all other methods achieved errors between 0.22s and 0.25s, and Pyannote was the worst with 0.62s (*Figure 5 and Supp. Table 4*). The threshold method showed 0.13s for offset error, followed by BreathGRU (0.15s), Fourier (0.16s), Silero (0.29s) and autoencoder (0.30s). BreathGRU demonstrated strong boundary localisation with the least onset and offset error for speech (onset = 0.24s, offset = 0.22s) and breath (onset = 0.10s, offset = 0.05s), indicating the best speech-breath transitions. Silero was competitive for breath but really performed poorly with speech durations and localisation. Fourier, autoencoder and Alam showed moderate errors, while Pyannote was the worst, with 0.62s for onset and 1.24s for offset, showing that these methods identify the regions correctly, but the boundaries were far from the true locations.

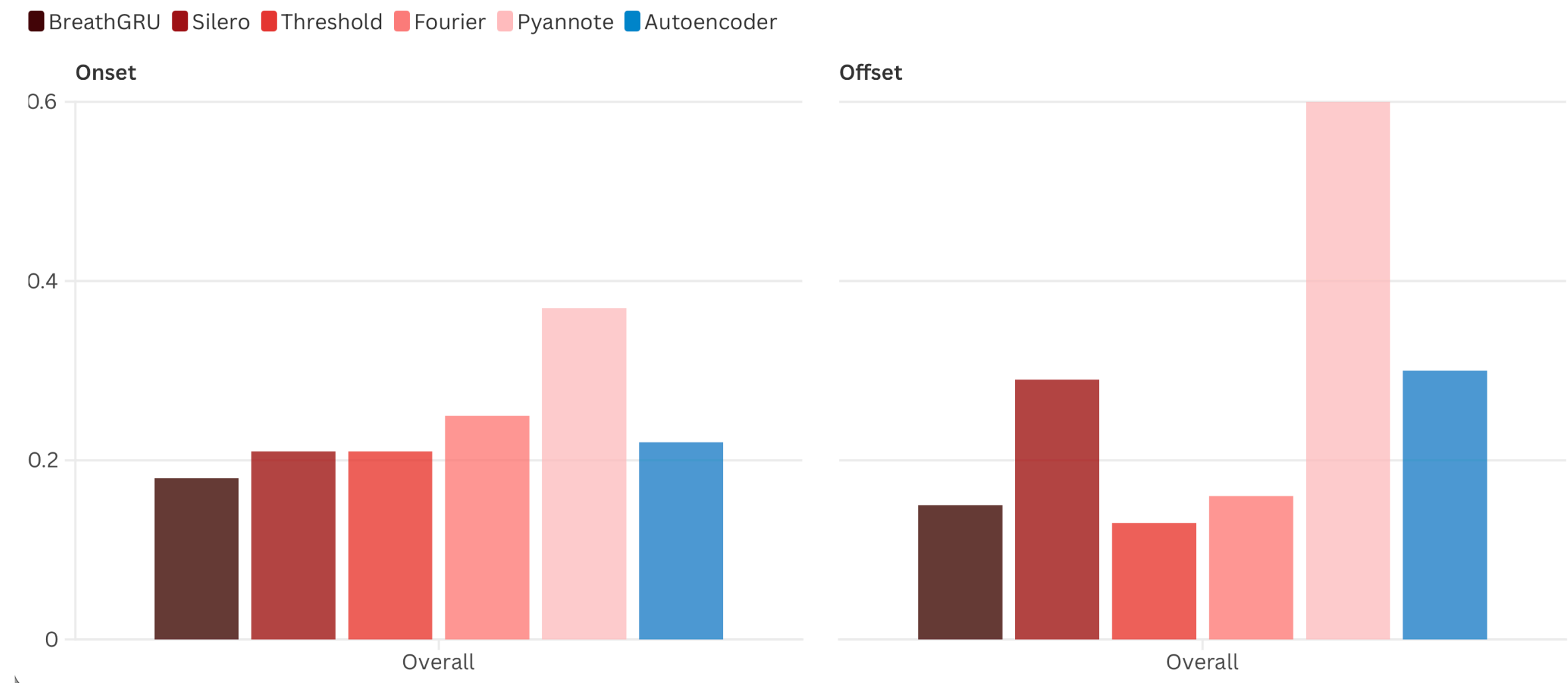

*Figure 5: The bar plot gives the onset and offset localisation in seconds. BreathGRU showed one of the lowest errors.*

**3.4. Time and Mean-Match IoU Evaluation:**

Figure 6 and Supp. Table 4 show the Time and Mean-Match IoU for the methods. Silero and BreathGRU achieve the highest temporal overlap (Time IoU) with scores of 0.72 and 0.63, respectively. For segmentation of speech and breath, Silero (0.89) and Pyannote (0.85) show the best speech temporal overlap as VADs, followed by BreathGRU (0.71). Breath overlap was almost similar between Silero (0.56) and BreathGRU (0.54), other methods showed variable values. Mean Match IoU showed the same score of 0.81 for Pyannote and BreathGRU, and 0.79 for Silero. The results suggest that BreathGRU achieves highly accurate segment overlap after identifying speech and breath segments.

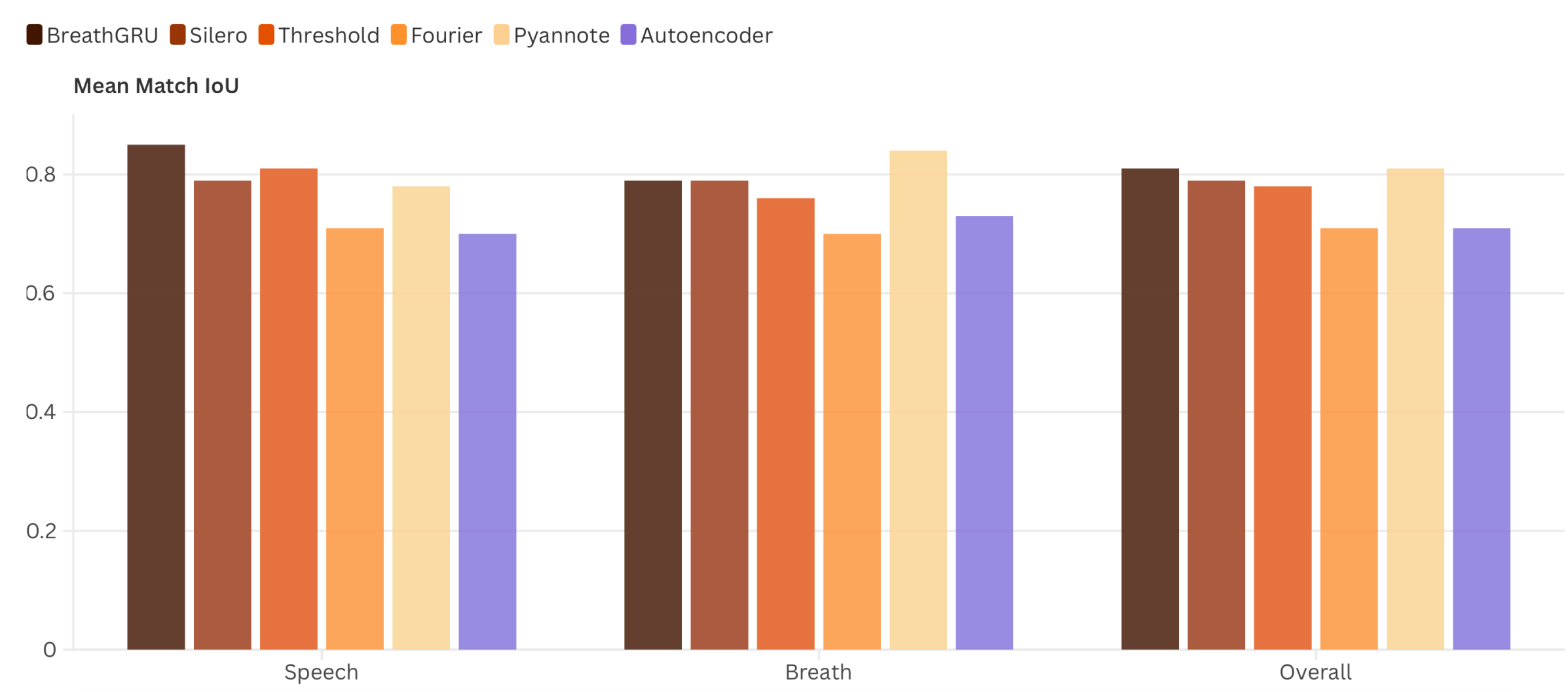

*Figure 6: This bar chart gives the Mean Match IoU for speech, breath and overall scores for all the models. BreathGRU displays the best performance for Mean Match IoU*

### 3.5. Duration-based Evaluation:

Silero was the best performing and closely matched the reference values with the least duration error of 2.41s (*Supp. Table 5*). BreathGRU and Pyannote had reasonably close mean duration errors of 4.89s and 7.11s, respectively. The signed duration error was calculated to determine whether a method over-segmented or under-segmented speech and breath. Silero showed signed duration bias of -1.81s and +1.81s, which indicates that it produces slightly shorter speech segments and slightly longer breath segments, similar to Pyannote and BreathGRU. While threshold and autoencoder models underestimated the speech durations and overestimated the breath durations, Fourier was the worst performing, with almost 22.21s of duration error.

### 3.6. Boundary Tolerance Evaluation:

The boundary tolerance again shows Silero as the best performing method, with 71% of the boundaries within the tolerance range. The threshold method was 63%, followed by BreathGRU (61%) and Pyannote (53%) and Autoencoder and Fourier were less than 40%. As for the speech boundaries, the threshold had the highest agreement with the ground truth at 65%, followed by BreathGRU (57%), Silero (50%), and the rest were less than 36%. Breath boundaries show a different trend, with Silero at 93%, Pyannote (68%), autoencoder (66%), BreathGRU (64%) and threshold (64%) showing similar scores, *see Supp. Table 7*.

### 3.7. Qualitative Comparison of Segmentation Performance:

*Figure 7* presents a 30-second recording from a patient in the dataset, which compares the proposed BreathGRU with manually annotated events and the pretrained Silero VAD model. Each outlined region corresponds to a breath event, showing a visual comparison across the three methods. BreathGRU closely follows the reference annotations for most of the breath events and produces consistent speech-breath transitions. In contrast, Silero fails to detect several breath events (highlighted in red) and partially detects many. Overall, the proposed framework demonstrates the closest agreement to the manual annotations and fewer false-positive breath events than the Silero model. The plots for other methods are in *Supp. Figure 2*.

### 3.8. Generalisation Testing on publicly available audio recordings:

The methods were tested for generalisation and performance on randomly selected fast and slow counting clips from Coswara datasets and TED Talks from YouTube. Each of the clips was visually analysed for segmentation, as ground-truth or manually extracted segmentation was not available. The recording displayed in the plots below is from an Indian actor's TED Talk, with an Indian accent and running speech. The segmentation in *Figure 8* shows a clear distinction in BreathGRU detecting speech and breath or pauses in the audio recording, compared to VADs like Silero and smoother transitions between speech and breath events in natural and accented speech.

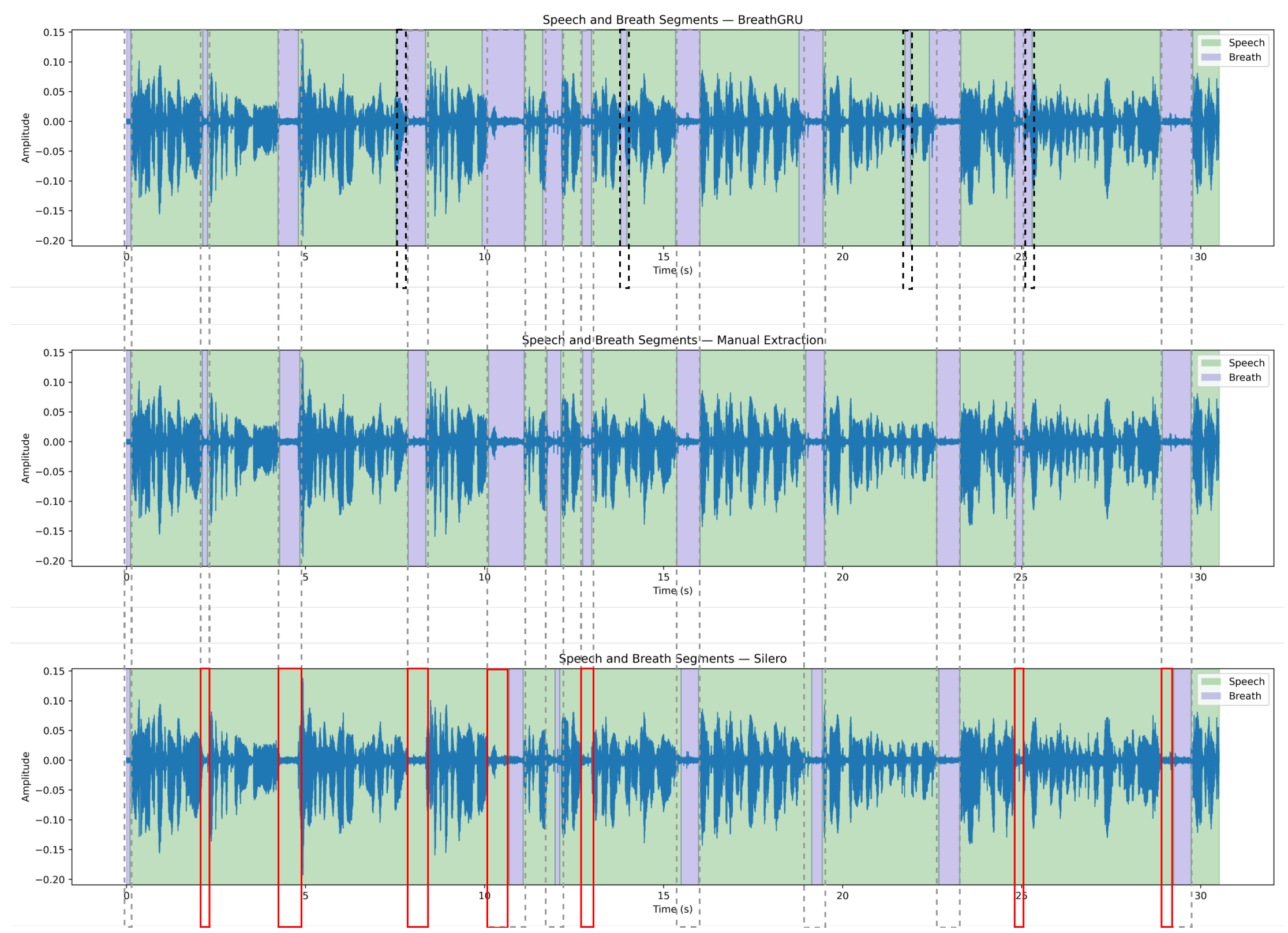

***Figure 7:*** *The frequency plots show speech and breath segmentation on a 30-second recording from BreathGRU, manual annotation, and Silero from one of the patient recordings. The highlighted segments show how BreathGRU accurately detects breath segments (grey dotted frames), whereas Silero misses major breath segments in the first two events. Red frames represent missed breath event segments. Dotted frames are false-positive events.*

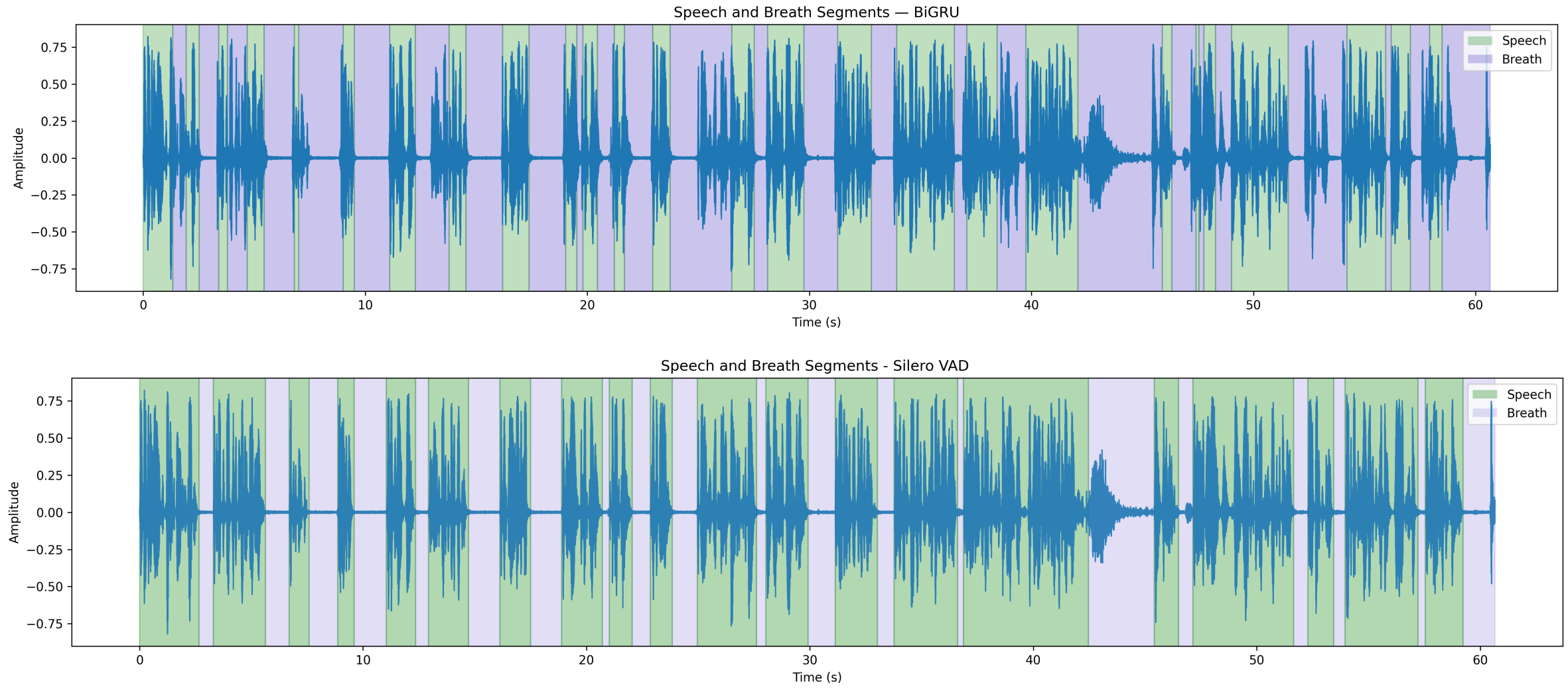

***Figure 8:*** *The frequency plots from 60 seconds of a TED Talk showing speech and breath segments as evaluated by the BiGRU and Silero models.*

# 4. Discussion

For applications such as lung function assessment using symptomatic sounds, audio preprocessing and segmentation are key steps. As speech and breath signals demonstrate differences in temporal and spectral features, it is essential to extract these two modalities for audio processing [5]. Inaccurate segmentation can affect the performance of these applications and predictive models. Existing approaches for speech-breath segmentation include the rule-based threshold method, the Fourier Transform method, unsupervised learning, and VAD approaches, but these methods fail to extract breath and wheeze, which are present in speech recordings in patients with chronic respiratory conditions.

This study proposed BreathGRU, a semi-supervised Bidirectional Gated Recurrent Unit framework for speech-breath segmentation and evaluated its performance against the existing models. Overall, the BreathGRU model demonstrated strong segmentation performance for multiple evaluation metrics while achieving the highest breath event recall and the lowest onset localisation error. These findings indicate the proposed framework is particularly effective for detecting respiratory events and accurately identifying speech-breath transitions, which are important for respiratory audio analysis.

One of the most notable findings was the better breath detection capability of BreathGRU, the method achieved the highest breath event recall (0.83), outperforming better than both Silero (0.67) and Pyannote (0.29). This suggests that BreathGRU is less likely to miss breath segments than VADs. Unlike pretrained VAD models, which are primarily trained to distinguish speech from non-speech, BreathGRU is trained specifically on annotated speech and breath recordings. Hence, the model learned acoustic representations associated with both speech and breathing while reducing the misclassification of respiratory sounds as silence or non-speech sounds. Since accurate breath sound detection is fundamental for respiratory health monitoring, this is an important advantage of the proposed framework.

BreathGRU also demonstrated strong temporal localisation performance, achieving the lowest onset localisation error and the highest Mean Match IoU (0.81), the second-highest Time-based F1 score (0.74), and the second-highest Time IoU (0.63). Although Silero achieved higher overall and duration-based metrics, BreathGRU more accurately identified the onset of breathing events and computed competitive segmentation performance. These results suggest that the proposed framework is effective at detecting temporal breath events and accurately localising speech-breath boundaries, both of which are important for respiratory feature extraction.

The observed metrics are likely due to the architectural design of the BreathGRU framework, where the bidirectional recurrent layers incorporate information from both preceding and succeeding acoustic frames, which enables the model to exploit temporal context when identifying transitions between speech and breath. In addition, the integration of log-normal duration priors and Segmental Viterbi decoding encourages consistent predictions by reducing rapid switching between speech-breath classes. This combination of bidirectional temporal modelling and duration-consistent decoding is responsible for the improved onset localisation and high Mean Match IoU achieved by the proposed framework.

An important observation is that BreathGRU achieved performance comparable to large pretrained VAD models despite being developed using a smaller set of recordings from asthmatic patients. Silero and Pyannote have been trained on thousands of hours of annotated speech recordings for speech activity detection, whereas BreathGRU was competitive for speech segmentation and superior for breath sound detection. These findings highlight the importance of breath-specific modelling, which provides an advantage over generic speech detection models.

The qualitative comparison presented in *Figure 7* further supports the quantitative findings when compared to manually annotated labels, BreathGRU accurately detects most breath events. In contrast, Silero fails to detect several breath segments, particularly in the early portion of the recordings and partially identifies others. Across the twelve manually annotated breath events, BreathGRU detects all events while producing two false-positive predictions, whereas Silero correctly detects only five breath events, completely misses five events and partially detects two. Although this example is a qualitative illustration, it demonstrates the improved breath detection capability of the proposed framework and visually supports the event-level performance of the model.

Accurate breath segmentation is particularly important for respiratory healthcare applications, including airflow assessment and extraction of respiratory acoustic biomarkers. For these applications, misclassifications of breath events may have clinical implications. The high breath event recall achieved by BreathGRU is especially relevant for respiratory analysis while preserving complete breathing cycles. The qualitative evaluation on unseen recordings from Coswara and publicly available YouTube recordings demonstrates the consistent segmentation performance beyond the training dataset. Although ground-truth annotations were not available, the proposed framework showed consistent speech-breath segmentation across diverse recording conditions and accents. These observations suggest that BreathGRU exhibits generalisation beyond the manually annotated respiratory recordings used during training, while further evaluation on larger external datasets is required to establish the robustness and clinical applicability of the framework.

## 5. Conclusion

This study proposed BreathGRU, a semi-supervised Bidirectional Gated Recurrent Unit (BiGRU) framework for speech-breath segmentation and evaluated it against a rule-based threshold method, a Fourier-based model, an unsupervised autoencoder model and pretrained Voice Activity Detection (VAD) models using manually annotated recordings. BreathGRU achieved the highest breath event recall, the lowest onset localisation error, and the highest Mean Match IoU while maintaining competitive overall segmentation performance.

These findings demonstrate that respiratory event modelling using bidirectional temporal context and duration-constrained decoding improves breath detection and boundary localisation compared with general-purpose VAD models. Future work will focus on integrating BreathGRU into respiratory health assessment pipelines for digital biomarker extraction, lung function prediction and respiratory disease monitoring.

**GitHub Repository:**
All code used in this study is publicly available at https://github.com/FaisalRezwan/BreathGRU.git. The repository contains test data and all scripts for data processing and analysis in this paper. The code is written in Python 3.9.6, and dependencies are specified in requirements.txt.

# Supplementary Material

**Metric Formulas:**

**vii. Event-based F1 Score:**

Event-based Precision:

$$Precision_{event} = \frac{TP}{TP + FP}$$

where, *TP* is the correctly detected events and *FP* is the false events.

Event-based Recall

$$Recall_{event} = \frac{TP}{TP + FN}$$

where, *FN* is the missed reference events.

Event-based F1 Score:

$$F1_{event} = \frac{2 \times Precision_{event} \times Recall_{event}}{Precision_{event} + Recall_{event}}$$

**viii. Time-based F1 Score**

Time-based Precision:

$$Precision_{time} = \frac{T_{overlap}}{T_{predicted}}$$

where, $T_{overlap}$ is the correctly classified duration and $T_{predicted}$ is the total predicted duration.

Time-based Recall:

$$Recall_{time} = \frac{T_{overlap}}{T_{referemce}}$$

where, $T_{reference}$ is the total reference duration.

Time-based F1 Score:

$$F1_{time} = \frac{2 \times Precision_{time} \times Recall_{time}}{Precision_{time} + Recall_{time}}$$

**ix. Onset and Offset Error:**

Mean Onset Error:

$$E_{onset} = \frac{1}{N}\sum_{i=1}^{N} |t_{pred,i}^{start} - t_{ref,i}^{start}|$$

Mean Offset Error:

$$E_{offset} = \frac{1}{N}\sum_{i=1}^{N} |t_{pred,i}^{end} - t_{ref,i}^{end}|$$

**x. Mean Duration Error**

$$E_{duration} = \frac{1}{N}\sum_{i=1}^{N} |D_{pred,i} - D_{ref,i}|$$

**xi. Mean Match Intersection Over Union (IoU):**

Intersection over Union (IoU):

For a predicted segment $S_{pred}$ and reference segment $S_{ref}$:

$$IoU = \frac{|S_{pred} \cap S_{ref}|}{|S_{pred} \cup S_{ref}|}$$

where, $|S_{pred} \cap S_{ref}|$ is the temporal overlap and $|S_{pred} \cup S_{ref}|$ is the combined duration covered by the segment

Mean Match IoU:

If $N$ matched segments exist:

$$Mean\ IoU = \frac{1}{N}\sum_{i=1}^{N} IoU_i$$

**xii. Boundary Tolerance**

For a tolerance threshold τ:

$$Boundary\ Tolerance = \frac{N_{within\ \tau}}{N_{total}}$$

where, $N_{within\ \tau}$ is the total number of predicted boundaries within tolerance and $N_{total}$ is the total number of evaluated boundaries.

*Supp. Table 1: The configurations of the STFT and Wiener-like mask used in the Fourier-based segmentation of speech and breath*

| Parameter | Value |
|---|---|
| FFT size | 2048 |
| Hop length | 512 |
| Speech score threshold | 0.55 |
| RMS threshold | 0.08 |
| Minimum speech duration | 0.12 s |
| Minimum breath duration | 0.08 s |

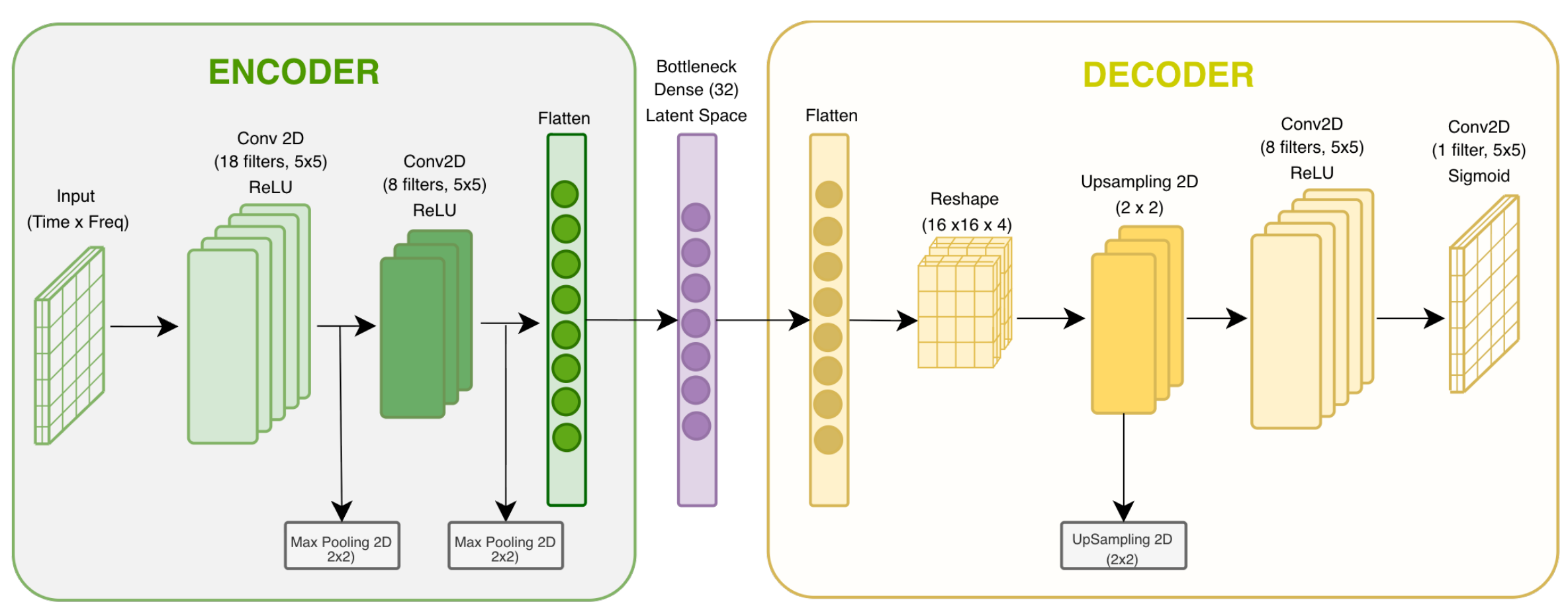

*Supp. Figure 1: The framework for the autoencoder with the complete encoder-decoder architecture.*

*Supp. Table 2: This table shows Event precision, recall and F1 scores for speech, breath and overall for each of the six models. Silero performs best at event precision, but BreathGRU is better at breath recall.*

| Model | Segment Type | Event Precision Mean | Event Precision std | Event Recall Mean | Event Recall std | Event F1 mean | Event F1 std |
|---|---|---|---|---|---|---|---|
| **threshold** | speech | 0.34 | 0.41 | 0.43 | 0.48 | 0.37 | 0.44 |
| **threshold** | breath | 0.44 | 0.37 | 0.60 | 0.41 | 0.50 | 0.38 |
| **threshold** | overall | 0.39 | 0.38 | 0.51 | 0.45 | 0.44 | 0.40 |
| **fourier** | speech | 0.02 | 0.03 | 0.05 | 0.07 | 0.03 | 0.04 |
| **fourier** | breath | 0.15 | 0.08 | 0.48 | 0.21 | 0.23 | 0.11 |
| **fourier** | overall | 0.09 | 0.09 | 0.27 | 0.27 | 0.13 | 0.13 |
| **silero** | speech | 0.54 | 0.26 | 0.69 | 0.31 | 0.59 | 0.28 |
| **silero** | breath | 0.51 | 0.15 | 0.67 | 0.29 | 0.55 | 0.20 |
| **silero** | overall | 0.53 | 0.21 | 0.68 | 0.29 | 0.57 | 0.24 |
| **pyannote** | speech | 0.43 | 0.37 | 0.25 | 0.31 | 0.30 | 0.34 |
| **pyannote** | breath | 0.61 | 0.33 | 0.29 | 0.22 | 0.36 | 0.22 |
| **pyannote** | overall | 0.52 | 0.35 | 0.27 | 0.27 | 0.33 | 0.28 |
| **autoencoder** | speech | 0.11 | 0.09 | 0.29 | 0.21 | 0.16 | 0.12 |
| **autoencoder** | breath | 0.05 | 0.08 | 0.13 | 0.22 | 0.08 | 0.12 |
| **autoencoder** | overall | 0.08 | 0.09 | 0.21 | 0.23 | 0.12 | 0.13 |
| **bigru** | speech | 0.38 | 0.39 | 0.51 | 0.45 | 0.43 | 0.41 |
| **bigru** | breath | 0.51 | 0.29 | 0.83 | 0.24 | 0.61 | 0.27 |
| **bigru** | overall | 0.45 | 0.34 | 0.67 | 0.38 | 0.52 | 0.35 |

*Supp. Table 3: The table presents mean precision, recall and F1 scores for Time-based segment detection for speech, breath and the combination of the two for all the models. Silero was the best-performing model for Time-based amongst the six methods.*

| Model | Segment Type | Time Precision Mean | Time Precision std | Time Recall Mean | Time Recall std | Time F1 Mean | Time F1 std |
|---|---|---|---|---|---|---|---|
| **threshold** | speech | 0.88 | 0.33 | 0.54 | 0.40 | 0.61 | 0.38 |
| **threshold** | breath | 0.49 | 0.29 | 0.95 | 0.05 | 0.60 | 0.25 |
| **threshold** | overall | 0.68 | 0.36 | 0.75 | 0.35 | 0.60 | 0.31 |
| **fourier** | speech | 0.99 | 0.01 | 0.34 | 0.15 | 0.49 | 0.18 |
| **fourier** | breath | 0.28 | 0.08 | 0.98 | 0.03 | 0.43 | 0.10 |
| **fourier** | overall | 0.63 | 0.37 | 0.66 | 0.35 | 0.46 | 0.14 |
| **silero** | speech | 0.93 | 0.04 | 0.96 | 0.04 | 0.94 | 0.02 |
| **silero** | breath | 0.83 | 0.16 | 0.65 | 0.26 | 0.69 | 0.23 |
| **silero** | overall | 0.88 | 0.13 | 0.80 | 0.24 | 0.81 | 0.20 |
| **pyannote** | speech | 0.87 | 0.04 | 0.98 | 0.03 | 0.92 | 0.02 |
| **pyannote** | breath | 0.84 | 0.22 | 0.36 | 0.24 | 0.46 | 0.25 |
| **pyannote** | overall | 0.86 | 0.16 | 0.67 | 0.36 | 0.69 | 0.29 |
| **autoencoder** | speech | 0.94 | 0.05 | 0.67 | 0.22 | 0.76 | 0.19 |
| **autoencoder** | breath | 0.12 | 0.13 | 0.39 | 0.45 | 0.17 | 0.19 |
| **autoencoder** | overall | 0.53 | 0.44 | 0.53 | 0.37 | 0.46 | 0.35 |
| **bigru** | speech | 0.99 | 0.01 | 0.72 | 0.25 | 0.81 | 0.18 |
| **bigru** | breath | 0.56 | 0.25 | 0.96 | 0.05 | 0.68 | 0.20 |
| **bigru** | overall | 0.78 | 0.28 | 0.84 | 0.22 | 0.74 | 0.19 |

*Supp. Table 4: The table gives the onset and offset localisation in seconds. Silero and BreathGRU showed competitive performances.*

| Model | Segment Type | Mean Onset Error Mean | Mean Onset Error std | Mean Offset Error Mean | Mean Offset Error std | Median Onset Error Mean | Mediano Onset Error std | Median Offset Error Mean | Median Offset Error std |
|---|---|---|---|---|---|---|---|---|---|
| **threshold** | speech | 0.28 | 0.23 | 0.19 | 0.07 | 0.11 | 0.19 | 0.09 | 0.05 |
| **threshold** | breath | 0.17 | 0.12 | 0.09 | 0.08 | 0.15 | 0.11 | 0.06 | 0.06 |
| **threshold** | overall | 0.21 | 0.18 | 0.13 | 0.09 | 0.13 | 0.14 | 0.08 | 0.06 |
| **fourier** | speech | 0.40 | 0.09 | 0.07 | 0.04 | 0.36 | 0.06 | 0.06 | 0.03 |
| **fourier** | breath | 0.18 | 0.14 | 0.21 | 0.13 | 0.14 | 0.12 | 0.13 | 0.14 |
| **fourier** | overall | 0.25 | 0.16 | 0.16 | 0.13 | 0.21 | 0.15 | 0.11 | 0.12 |
| **silero** | speech | 0.31 | 0.21 | 0.53 | 0.54 | 0.04 | 0.01 | 0.30 | 0.54 |
| **silero** | breath | 0.10 | 0.05 | 0.05 | 0.04 | 0.08 | 0.03 | 0.03 | 0.01 |
| **silero** | overall | 0.21 | 0.19 | 0.29 | 0.45 | 0.06 | 0.03 | 0.17 | 0.40 |
| **pyannote** | speech | 0.62 | 1.12 | 1.24 | 1.64 | 0.60 | 1.13 | 1.09 | 1.71 |
| **pyannote** | breath | 0.15 | 0.08 | 0.04 | 0.03 | 0.13 | 0.09 | 0.03 | 0.02 |
| **pyannote** | overall | 0.37 | 0.78 | 0.60 | 1.24 | 0.35 | 0.78 | 0.53 | 1.25 |
| **autoencoder** | speech | 0.26 | 0.14 | 0.41 | 0.25 | 0.17 | 0.09 | 0.28 | 0.29 |
| **autoencoder** | breath | 0.16 | 0.11 | 0.12 | 0.03 | 0.15 | 0.11 | 0.10 | 0.03 |
| **autoencoder** | overall | 0.22 | 0.13 | 0.30 | 0.24 | 0.16 | 0.10 | 0.21 | 0.24 |
| **bigru** | speech | 0.24 | 0.20 | 0.22 | 0.11 | 0.11 | 0.21 | 0.10 | 0.07 |
| **bigru** | breath | 0.14 | 0.08 | 0.10 | 0.09 | 0.10 | 0.08 | 0.07 | 0.09 |
| **bigru** | overall | 0.18 | 0.14 | 0.15 | 0.11 | 0.10 | 0.14 | 0.08 | 0.08 |

*Supp. Table 5: This table gives the Time IoU and Mean Match IoU for speech, breath and overall scores for all the models. Silero demonstrates the best performance for speech and breath Time IoU of 0.89, while BreathGRU displays the best performance for Mean Match IoU*

| Model | Segment Type | Time IoU Mean | Time IoU std | Mean Match IoU mean | Mean Match IoU std |
|---|---|---|---|---|---|
| **threshold** | speech | 0.53 | 0.39 | 0.81 | 0.17 |
| **threshold** | breath | 0.47 | 0.26 | 0.76 | 0.11 |
| **threshold** | overall | 0.50 | 0.33 | 0.78 | 0.13 |
| **fourier** | speech | 0.34 | 0.15 | 0.71 | 0.06 |
| **fourier** | breath | 0.28 | 0.08 | 0.70 | 0.10 |
| **fourier** | overall | 0.31 | 0.12 | 0.71 | 0.09 |
| **silero** | speech | 0.89 | 0.03 | 0.79 | 0.11 |
| **silero** | breath | 0.56 | 0.23 | 0.79 | 0.10 |
| **silero** | overall | 0.72 | 0.23 | 0.79 | 0.10 |
| **pyannote** | speech | 0.85 | 0.04 | 0.78 | 0.17 |
| **pyannote** | breath | 0.33 | 0.23 | 0.84 | 0.08 |
| **pyannote** | overall | 0.59 | 0.31 | 0.81 | 0.13 |
| **autoencoder** | speech | 0.63 | 0.20 | 0.70 | 0.07 |
| **autoencoder** | breath | 0.10 | 0.12 | 0.73 | 0.03 |
| **autoencoder** | overall | 0.37 | 0.32 | 0.71 | 0.06 |
| **bigru** | speech | 0.71 | 0.25 | 0.85 | 0.09 |
| **bigru** | breath | 0.54 | 0.23 | 0.79 | 0.08 |
| **bigru** | overall | 0.63 | 0.25 | 0.81 | 0.09 |

*Supp. Table 6: The table gives absolute and signed error for speech, breath and average scores for all six models. Silero has the least absolute error in detecting duration boundaries.*

| Model | Segment Type | Duration abs error mean | Duration abs error std | Duration Signed Error Mean | Duration Signed Error std |
|---|---|---|---|---|---|
| **threshold** | speech | 11.24 | 9.51 | -11.24 | 9.51 |
| **threshold** | breath | 11.24 | 9.51 | 11.24 | 9.51 |
| **threshold** | overall | 11.24 | 9.22 | 0.00 | 14.80 |
| **fourier** | speech | 22.38 | 7.51 | -22.38 | 7.51 |
| **fourier** | breath | 22.05 | 7.26 | 22.05 | 7.26 |
| **fourier** | overall | 22.21 | 7.17 | -0.17 | 23.96 |
| **silero** | speech | 2.41 | 2.57 | 1.81 | 3.08 |
| **silero** | breath | 2.41 | 2.58 | -1.81 | 3.08 |
| **silero** | overall | 2.41 | 2.50 | 0.00 | 3.52 |
| **pyannote** | speech | 4.89 | 3.57 | 4.89 | 3.57 |
| **pyannote** | breath | 4.88 | 3.56 | -4.88 | 3.56 |
| **pyannote** | overall | 4.89 | 3.46 | 0.00 | 6.10 |
| **autoencoder** | speech | 9.72 | 8.13 | -9.69 | 8.17 |
| **autoencoder** | breath | 16.06 | 15.51 | 11.67 | 19.43 |
| **autoencoder** | overall | 12.89 | 12.45 | 0.99 | 18.16 |
| **bigru** | speech | 7.11 | 5.75 | -7.11 | 5.75 |
| **bigru** | breath | 7.12 | 5.75 | 7.12 | 5.75 |
| **bigru** | overall | 7.11 | 5.58 | 0.00 | 9.20 |

*Supp. Table 7: The table gives scores for boundary tolerance in detecting speech and breath segments. Silero shows that 71% of the segments match the ground truth.*

| Model | Segment Type | Boundaries within tolerance mean | Boundaries within tolerance std |
|---|---|---|---|
| **threshold** | speech | 0.65 | 0.38 |
| **threshold** | breath | 0.62 | 0.42 |
| **threshold** | overall | 0.63 | 0.38 |
| **fourier** | speech | 0.23 | 0.26 |
| **fourier** | breath | 0.42 | 0.37 |
| **fourier** | overall | 0.36 | 0.34 |
| **silero** | speech | 0.50 | 0.24 |
| **silero** | breath | 0.93 | 0.11 |
| **silero** | overall | 0.71 | 0.28 |
| **pyannote** | speech | 0.36 | 0.39 |
| **pyannote** | breath | 0.68 | 0.24 |
| **pyannote** | overall | 0.53 | 0.35 |
| **autoencoder** | speech | 0.22 | 0.19 |
| **autoencoder** | breath | 0.66 | 0.24 |
| **autoencoder** | overall | 0.38 | 0.29 |
| **bigru** | speech | 0.57 | 0.33 |
| **bigru** | breath | 0.64 | 0.31 |
| **bigru** | overall | 0.61 | 0.31 |

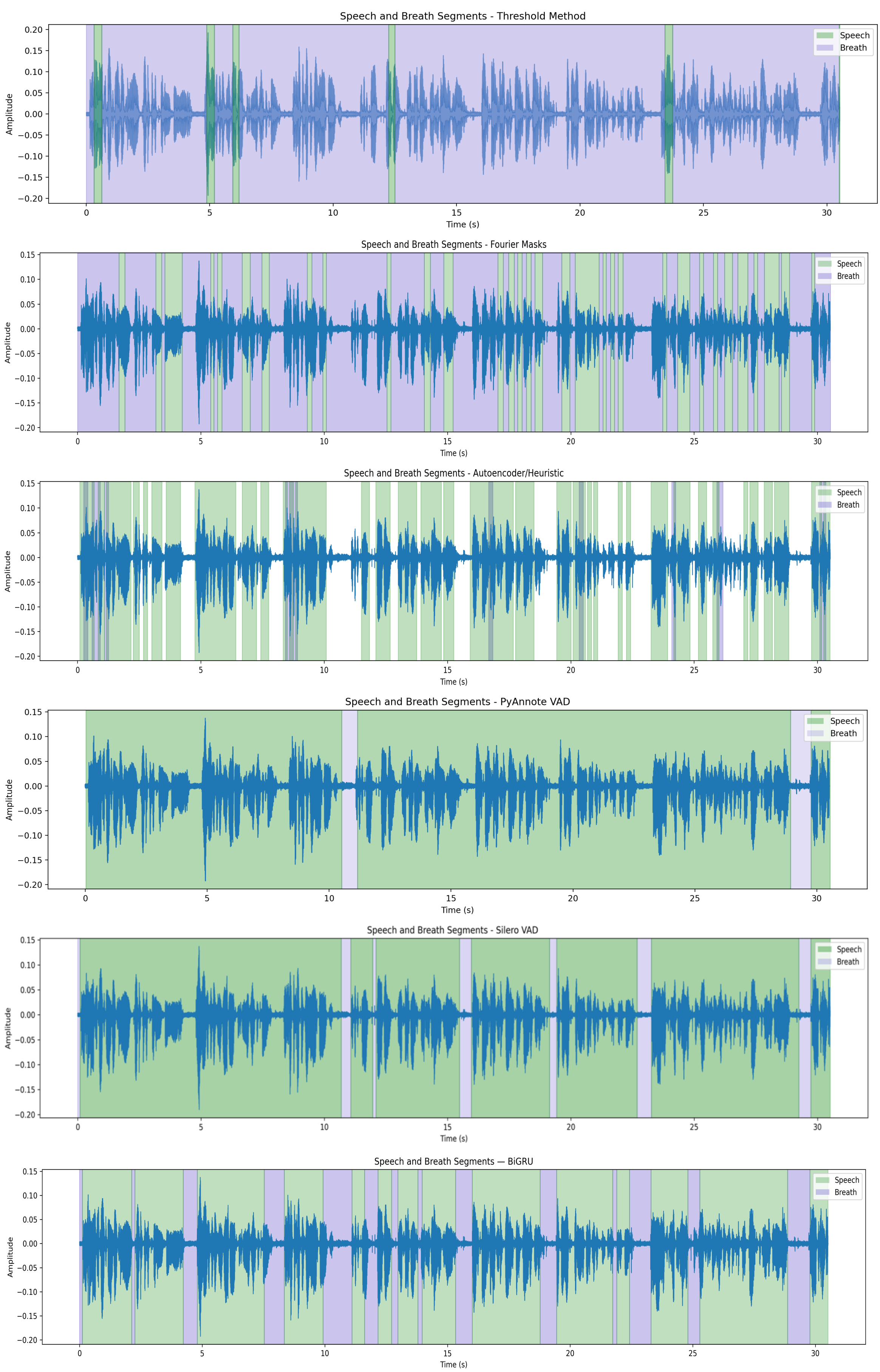

***Supp. Figure 2: The six frequency plots show speech and breath segmentation on a 30-second recording from the models***